\documentclass[10pt,aps,prl,twocolumn,showpacs,amsmath,superscriptaddress]{revtex4-1}

\usepackage{amsmath}
\usepackage{amsfonts}
\usepackage{amssymb}
\usepackage{mathrsfs}
\usepackage{graphicx}
\usepackage{verbatim}
\usepackage{bm}
\usepackage{multirow}
\usepackage[colorlinks,linkcolor=blue,citecolor=blue,urlcolor=blue]{hyperref}

\renewcommand{\i}{\mathrm{i}}
\newcommand{\e}{\mathrm{e}}

\renewcommand{\vec}[1]{\bm{#1}}
\newcommand{\algb}[1]{\mathfrak{#1}}

\begin{document}
	\title{Transformed Composite Sequences for Improved Qubit Addressing}
	\author{J. True Merrill}
	\email{true.merrill@gtri.gatech.edu}
	\affiliation{Georgia Tech Research Institute, Atlanta, GA 30332}
	\affiliation{School of Chemistry and Biochemistry, Georgia Institute of Technology, Atlanta, GA 30332}
	\author{S. Charles Doret}
	\altaffiliation[Current address: ]{Department of Physics, Harvey Mudd College, Claremont, CA 91711}
	\affiliation{Georgia Tech Research Institute, Atlanta, GA 30332}
	\author{Grahame D. Vittorini}
	\altaffiliation[Current address: ]{Joint Quantum Institute and Department of Physics, University of Maryland, College Park, MD 20742}
	\affiliation{School of Physics, Georgia Institute of Technology, Atlanta, GA 30332}
	\author{J.P. Addison}
	\affiliation{School of Physics, Georgia Institute of Technology, Atlanta, GA 30332}
	\author{Kenneth R. Brown}
	\affiliation{School of Chemistry and Biochemistry, Georgia Institute of Technology, Atlanta, GA 30332}
	\affiliation{School of Physics, Georgia Institute of Technology, Atlanta, GA 30332}
	\affiliation{School of Computational Science and Engineering, Georgia Institute of Technology, Atlanta, GA 30332}
	\date{\today}

\begin{abstract}
Selective laser addressing of a single atom or atomic ion qubit can be improved using narrowband composite pulse sequences.  We describe a Lie-algebraic technique to generalize known narrowband sequences and introduce new sequences related by dilation and rotation of sequence generators. Our method improves known narrowband sequences by decreasing both the pulse time and the residual error.  Finally, we experimentally demonstrate these composite sequences using $^{40}$Ca$^+$ ions trapped in a surface-electrode ion trap.  
\end{abstract}


\maketitle

In ion trap quantum computers~\cite{Cirac1995, Nagerl1999} and neutral atom optical lattices~\cite{Scheunemann2000, Dumke2002}, single-qubit addressing typically requires focused lasers where the beam waist is smaller than the inter-atom separation.  Closely spaced atoms are generally desirable to improve two-qubit coupling rates, often demanding inter-atom spacings approaching the diffraction limit.  In practice, single-qubit addressing requires precise focal alignment and ultra-stable beam steering to prevent unwanted excitations on neighboring atoms, a significant challenge as the number of qubits increases~\cite{Knoernschild2010}.  Furthermore, achieving the tight focus required for single-ion addressing is often made difficult by geometric constraints and restricted optical access ~\cite{Seidelin2006, Doret2012}.  These factors combine to make single-qubit addressing a major challenge in many experimental systems.

One method of improving single-qubit addressing applies an auxiliary field gradient to shift qubit transition frequencies, affording a degree of selective control~\cite{Wang2009, Zhang2006}.  A similar technique uses an intense laser to introduce a position-dependent AC Stark shift~\cite{Staanum2002,  Weitenberg2011}. Recently, quantum control has been used in conjunction with frequency shifts to achieve addressing with inhomogeneous control fields \cite{Jessen2013}. A recent proposal has also examined spatial refocusing through precise laser positioning coupled with controlled phase shifts \cite{ShenPRA2013}.  These methods frequently require time-consuming calibrations to remove systematic errors while adding to experimental complexity, limiting scalability.  These techniques also make strong physical assumptions of the nature of the qubit, and are generally not translatable to other qubit technologies.  

In this article we demonstrate an alternative control scheme that replaces simple single-qubit gates with a narrowband composite sequence of laser pulses designed for local addressing~\cite{Ivanov2011, Merrill2013}.  These sequences allow the exclusive manipulation of a single qubit even when neighboring qubits are subjected to significant laser intensity.   Such compensating sequences reduce systematic control errors at the cost of increasing the time required to produce gates~\cite{Merrill2013}.   
Our main result is a new technique to generate fully-compensating narrowband sequences using Lie-algebraic transformations of other known sequences.  We use numerical minimization to identify sequences with superior error correction properties and low operation times compared to the original sequence family.   Further, we demonstrate the effectiveness of these sequences for single-qubit addressing in an experiment with $^{40}$Ca$^+$ ion qubits in a surface-electrode trap.

Here we consider a register of $N$ identical spatially separated qubits.  A resonant laser in the rotating-wave limit illuminates an addressed qubit $i$, but also illuminates neighboring qubits $j$ at a lower intensity ($j \neq i$), resulting in an addressing error that yields a separable but correlated evolution on each qubit.  Control over the qubits is implemented by applying a time-dependent Hamiltonian
\begin{align}
H(t) = \frac{\hbar \Omega_i}{2} \left\{ \sigma_i(\varphi(t)) +\sum_ {j\neq i} \epsilon_j \sigma_j(\varphi(t)) \right\},
\label{eq:hamiltonian}
\end{align}
where $\varphi(t)$ is the laser phase, $\Omega_i$ is the Rabi frequency for the addressed qubit $i$, and $\sigma(\varphi(t)) = X \cos \varphi(t) + Y \sin \varphi(t)$, where $X$, $Y$ are Pauli operators.  For simplicity we fix $|\Omega_i|^2$ to some maximal bounded value corresponding to the intensity peak of the laser field.  The terms $\epsilon_j \sigma_j(\varphi(t))$ induce undesired correlated rotations on neighboring qubits.  Here the ratio $\epsilon_j = \Omega_j / \Omega_i < 1$, where $\Omega_j$ is the Rabi frequency at the neighboring qubit $j$.  The frequency $\Omega_j$ parameterizes the magnitude of the addressing error and is assumed to be fixed over the entire duration of the control.  The time dependence of $H(t)$ is entirely due to the temporal modulation of the phase $\varphi(t)$, which here serves as our only control parameter.  This choice confers no loss of generality, since solutions with a time-dependent laser intensity may be considered with an appropriate substitution of the time variable. 

Compensating pulse sequences choose a control trajectory $\varphi(t)$ to yield a net evolution robust against a particular class of systematic errors.  A common simplification for $\varphi(t)$ is to divide the time coordinate into $L$ time intervals $(\Delta t_1, \Delta t_2, \dots, \Delta t_L)$ for which the phase is a constant angle $(\varphi_1, \varphi_2, \dots, \varphi_L)$.  Each pulse applies a spin rotation controlled by the generator of rotations $r_\ell = -\i \theta_\ell \sigma(\varphi_\ell) / 2$, where $\theta_\ell = \Omega_i \Delta t_\ell$ is the pulse area or rotation angle applied to the addressed qubit.  The total propagator for the entire sequence is $U(\vec{r}) = U_i(\vec{r}) \left[ \bigotimes_{j \neq i} U_j(\vec{r}) \right]$ where 
\begin{align}
U_i(\vec{r}) = \prod_{\ell=1}^L \exp \{ r_\ell \}, \qquad U_j(\vec{r}) = \prod_{\ell=1}^L \exp \{ \epsilon_j r_\ell \}
\end{align}
are the gates applied to the addressed qubit $i$ and the neighbor qubit $j$ respectively, and $\vec{r} = (r_{1}, r_{2}, \dots, r_{L})$ is the ordered set of rotation generators.  Following the usual convention, multiplication of successive pulse propagators occurs from the left to ensure the correct time-ordering.  With a careful choice of rotation generators, it is possible to produce propagators that apply a nontrivial gate $U_i(\vec{r}) = U_T$ to the addressed qubit while simultaneously approximating the identity $U_j(\vec{r}) = I + O(\epsilon_j^{n + 1})$ on all neighboring qubits.  Sequences with this property are called $n$th-order fully-compensating narrowband sequences \cite{Brown2004, Ivanov2011, Merrill2013, Low2013}.  So long as $\Omega_j \ll \Omega_i$ these sequences improve local gates on a addressed qubit even if the laser simultaneously illuminates several qubits.

We remark that an $n$th-order narrowband sequence must satisfy a set of $n$ Lie-algebraic constraints on the rotation generators.  Applying the Baker-Campbell-Hausdorff lemma we find that $U_j(\vec{r}) = \exp \left \{ \sum_{m = 1}^{\infty} \epsilon^m_j {F}_m (\vec{r}) \right \}$, where $F_m (\vec{r})$ is given by the generators and their commutators and is related to the $m$th-order average Hamiltonian.  Explicitly the first two terms are 
\begin{align}
	F_1(\vec{r}) = \sum_{\ell = 1}^{L} r_{\ell}, \qquad
	F_2(\vec{r}) = \frac{1}{2} \sum_{\ell = 1}^{L}\sum_{k = 1}^{\ell} [ r_{\ell}, r_{k} ]. 
\end{align}
To satisfy $U_j = I + O(\epsilon_j^{n+1})$ for all values of $\epsilon_j$, each $F_m(\vec{r})$ with $m \leq n$ must independently equal zero.  Frequently it is possible to assign a geometric interpretation to each constraint.  For example, $F_1(\vec{r}) = 0$ requires the generators $\vec{r}$ to form a closed figure on the Lie algebra, and $F_2(\vec{r}) = 0$ requires that the figure encloses signed areas of equal magnitude but opposite sign.  

\begin{figure}
  \begin{center}
    \includegraphics{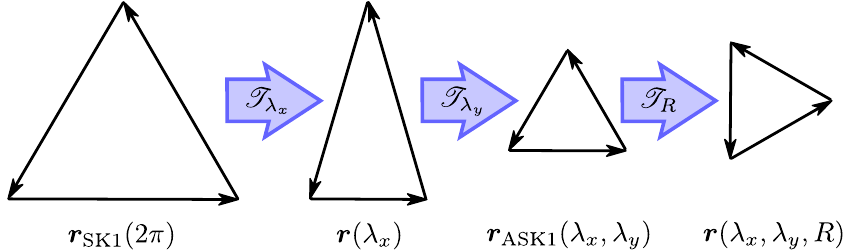}
  \end{center}
  \caption{Construction of $\vec{r}_{\text{ASK1}}(\lambda_x, \lambda_y)$ by compositions of dilation and rotation maps applied to $\vec{r}_{\text{SK1}}(2\pi)$. The maps ensure $F_1(\vec{r}_{\text{ASK1}}(\lambda_x, \lambda_y)) = 0$ and therefore $\vec{r}_{\text{ASK1}}(\lambda_x, \lambda_y)$ generates a first-order narrowband sequence.  Particular choices of $(\lambda_x, \lambda_y, R)$ result in different net rotations, total sequence pulse areas, and residual sequence errors. 
  }
    \label{figure1}
\end{figure}

We introduce a method to generalize existing narrowband sequences by identifying Lie-algebraic transformations on the generators which leave the constraint equations satisfied.  These transformations yield derivative sequences which achieve the same order of error suppression, but may offer substantial improvements in the total composite gate time as well as the gate accuracy.  Our method can be described as follows.  Let $\mathscr{T}_\lambda : \algb{su}(2) \mapsto \algb{su}(2)$ be a map between Lie algebra elements with the condition that if $F_m(\vec{r}) = 0$, then $F_m ( \mathscr{T}_\lambda \circ \vec{r}) = 0$  for all $m \leq n$.  This ensures that if $\vec{r}$ generates an $n$th-order compensating sequence, then $\vec{r}(\lambda) = \mathscr{T}_\lambda \circ \vec{r}$ also generates a sequence of the same order, however in general $U_i(\vec{r}(\lambda)) \neq U_i(\vec{r})$.  To find sequences that implement a particular target gate $U_T$, we perform an optimization over the mapped sequences to minimize a cost functional while constraining $U_i(\vec{r}(\lambda)) = U_T$.  Two cost functionals we consider are the total pulse area $\theta_\text{Total} = \sum_\ell^L |\theta_\ell|$, related to the total time required to perform a composite gate, and the infidelity $\mathcal{I} = 1 - \mathcal{F}$, where $\mathcal{F} = |\text{tr}[U_j(\vec{r}(\lambda))]/2|$ is the fidelity of the effective identity gate on the neighboring qubits. 

Maps that satisfy the constraint condition are common affine transformations. For arbitrary sequences, compositions of rotations and dilations fulfill the requirement: $F_m(R \vec{r} R^\dagger) = R F_m(\vec{r}) R^\dagger$ and $F_m(\lambda\vec{r}) = \lambda^m F_m(\vec{r})$. Independent dilation of each axis $\mathscr{T}_{\lambda_x} \circ X = \lambda_x X$ and $\mathscr{T}_{\lambda_y} \circ Y = \lambda_y Y$ will also satisfy this criteria for $n \leq 2$, and in our case, where the controls are restricted to the $X$-$Y$ plane, for $n \leq 3$.  Starting with an initial seed sequence $\vec{r}$ we generate a family of related sequences $\vec{r}(\lambda_x, \lambda_y, R)$ by the composition of dilations and rotations (see Fig.~\ref{figure1}).  

As an example, consider the first-order passband SK1 pulse sequences, produced by the generators $\vec{r}_\text{SK1}(\theta)  = (-\i \theta \sigma(0)/2 , -\i \pi \sigma(\varphi_\text{SK1}), -\i \pi \sigma(-\varphi_\text{SK1}))$ where $\cos \varphi_\text{SK1} =\theta/(4\pi)$~\cite{Brown2004}.  On the addressed qubit SK1 applies $U_i(\vec{r}_{\text{SK1}}(\theta)) = \exp(- \i \theta X / 2)$ and it approximates the identity on neighboring qubits, $U_j(\vec{r}_{\text{SK1}}(\theta))=I+O(\epsilon_j^2)$.  To illustrate our transformation method, we start with the specific case $\vec{r}_{\mathrm{SK1}}(2\pi)$ and identify a map which recovers the entire SK1 family.  Let $\mathscr{T}_\theta$ be the one parameter map that contracts the $X$ components by $\lambda_x=\theta/(2\pi)$ and expands the $Y$ components by $\lambda_y=\sqrt{(4-\lambda_x^2)/3}$.  This map satisfies $F_1(\mathscr{T}_\theta \circ \vec{r}_{\mathrm{SK1}}(2\pi)) = 0$ and $\vec{r}_{\mathrm{SK1}}(\theta) = \mathscr{T}_\theta \circ \vec{r}_{\mathrm{SK1}}(2\pi)$.   

SK1 can implement an arbitrary single-qubit gate using extra rotations,  $U_T = R U_i(\vec{r}_{\text{SK1}}(\theta)) R^\dagger$.  Alternatively, one simply changes the sequence generators using the similarity transformation $\vec{r}_{\text{SK1}} (\theta, R) = R \vec{r}_{\text{SK1}}(\theta) R^\dagger$. For a target in-plane rotation $U_T=\exp(- \i \theta \sigma(\varphi_T) / 2)$ this can be achieved by advancing all the $\varphi_l$ in SK1 by $\varphi_T$.

The composition of independent $X$ and $Y$ dilation maps applied to $\vec{r}_{\mathrm{SK1}}(2\pi)$ generates a larger family of narrowband sequences that we call ASK1, $\vec{r}_{\text{ASK1}}(\lambda_x, \lambda_y) = \mathscr{T}_{\lambda_y} \circ \mathscr{T}_{\lambda_x} \circ \vec{r}_\text{SK1} (2 \pi)$.  Fig.~\ref{figure1} illustrates the construction of ASK1 sequences.   Note that ASK1 usually applies a net rotation $U_i(\vec{r}_{\text{ASK1}}(\lambda_x, \lambda_y))$ about an axis outside of the $X$-$Y$ plane; such a sequence cannot replace an in-plane rotation implemented by a single pulse with a constant phase.   To achieve a target in-plane gate $U_T$ using the Hamiltonian \eqref{eq:hamiltonian} and phase advances, we introduce the similarity transformation $\vec{r}_{\text{ASK1}} (\lambda_x, \lambda_y, R) = R \vec{r}_{\text{ASK1}}(\lambda_x, \lambda_y) R^\dagger$, where $U_T = U_i(\vec{r}_{\text{ASK1}}(\lambda_x, \lambda_y, R))$.  We decompose $R = \exp( r' ) T$, where $\exp( r' )$ applies the minimum-angle rotation to match the polar angle of the rotation axis; $T$ is a rotation about $Z$ which can be implemented by a uniform phase advance on the inner ASK1 pulses.  The phase advance controls the azimuthal angle of the net rotation axis.  This transformed sequence construction, which we call TASK1, sets the net rotation angle with the innermost pulses.  In terms of rotation generators the sequence is $\vec{r}_{\text{TASK1}}(\lambda_x, \lambda_y, R) = (r', T \vec{r}_{\text{ASK1}}(\lambda_x, \lambda_y) T^\dagger, -r')$.     

\begin{figure}
  \begin{center}
    \includegraphics{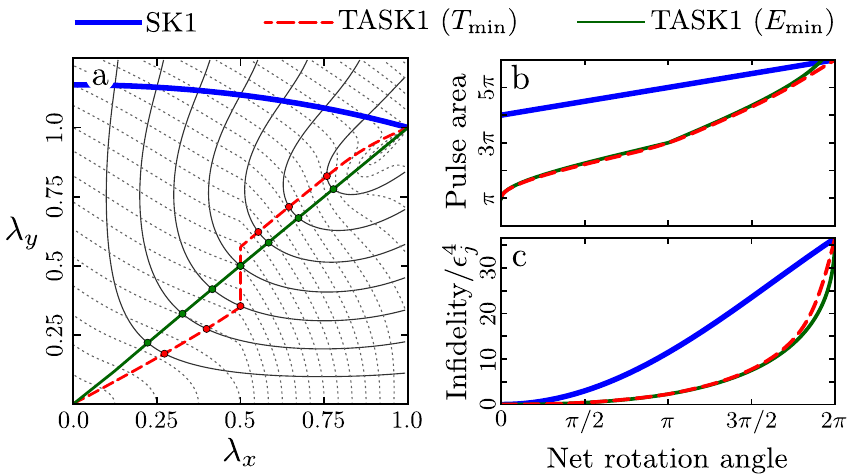}
  \end{center}
  \caption{ (a) The narrowband TASK1 family in terms of the scale parameters $(\lambda_x,\lambda_y)$.  Each TASK1 sequence implements a net unitary $U_T = \exp(-\i \theta X / 2)$ on addressed qubits, $\theta$ is the net rotation angle.  Contours of $\theta$ (solid) and the total pulse area (dashed) are plotted in intervals of $\pi/4$.  The SK1 sequences (blue, thick solid) are a subfamily of TASK1. (b) TASK1~($T_\text{min}$) (red, dashed) is the subfamily which minimizes the total pulse area.  (c) TASK1~($E_\text{min}$) (green, solid) minimizes the infidelity.  The time-minimal and error-minimal subfamilies outperform SK1 in both speed and accuracy. 
  }
  \label{figure2}
\end{figure}

Despite the inclusion of two additional pulses, the TASK1 sequences outperform SK1 in both the total pulse area and the infidelity.  
 Fig. \ref{figure2}a shows the TASK1 family in terms of $(\lambda_x, \lambda_y)$ and plots contours of the net rotation angle and total pulse area.  
 Using constrained optimization we identify a subfamily of sequences that minimize the total pulse area, TASK1~($T_\text{min}$), and the infidelity, TASK1~($E_\text{min}$) for a fixed net rotation angle.  We find the error-minimal sequences correspond to $\lambda_x = \lambda_y$ and result in ASK1 sequences homologous to equilateral triangles in  $\algb{su}(2)$. Fig. \ref{figure2}b compares the infidelity and total pulse area for each sequence subfamily.  We see that TASK1~($T_\text{min}$) and TASK1~($E_\text{min}$) outperform SK1 in both the required time and the minimization of the residual rotation on the neighboring qubit.  We also note that the time minimal and error minimal sequences yield similar performance for most net rotation angles.  In particular, for a target rotation $U_T = \exp \left( -\i \pi X / 2 \right)$, the error-minimal and time-minimal sequences are identical, $\lambda_x = \lambda_y = 1/2$, and TASK1 performs the gate using $3/5$ of the total pulse-area and with $1/5$ of the residual infidelity compared to SK1.  Explicit descriptions of the pulses can be found in the Supplementary Material.

\begin{figure}
  \begin{center}
    \includegraphics[width=0.8\columnwidth]{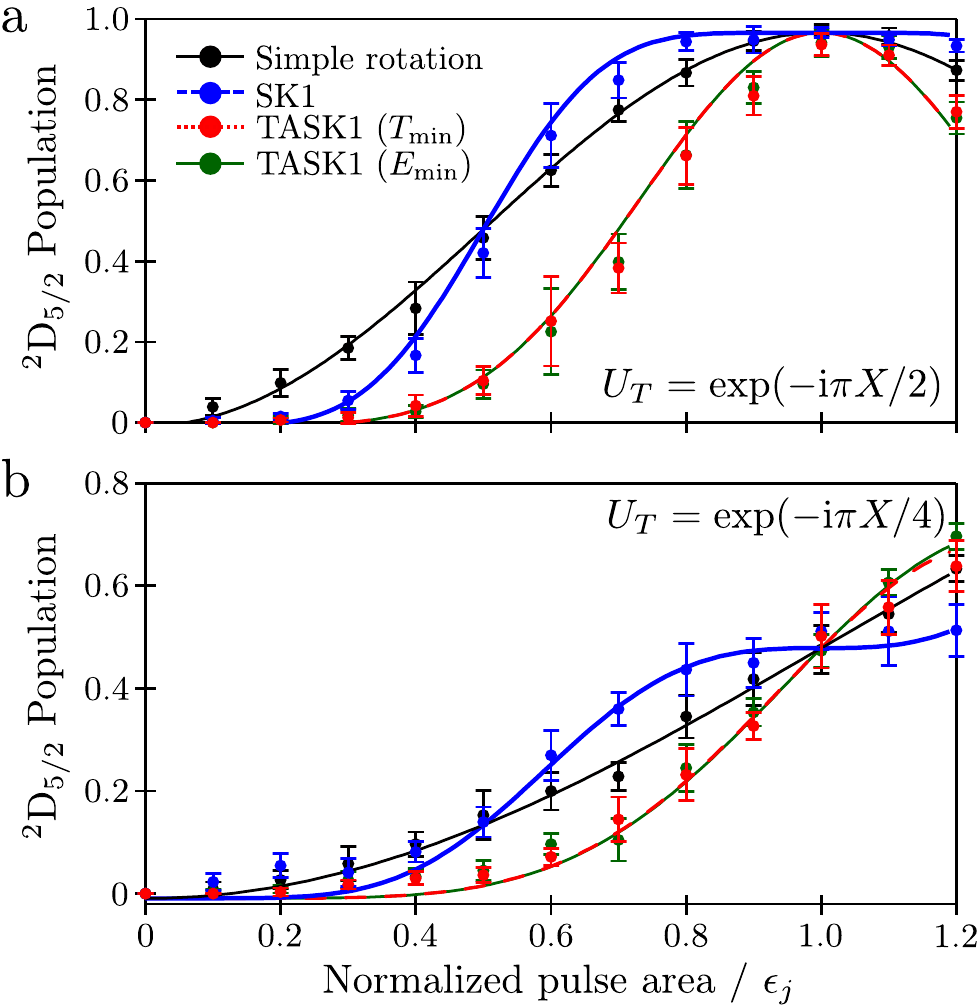}
  \end{center}
  \caption{Population inversion as a function of normalized pulse area $\epsilon_j$ for target gates $U_T$ composed of simple rotations or by SK1, TASK1~($T_\text{min}$), or TASK1~($E_\text{min}$) sequences.  In (a) $U_T = \exp(-\i \pi X/2)$ and in (b) $U_T = \exp(-\i \pi X/4)$.  Curves are theoretical predictions with a single, common adjustable parameter accounting for experimental qubit detection fidelity.  Narrowband sequences suppress population inversion for small $\epsilon_j$ but perform $U_T$ when $\epsilon_j = 1$. 
    \label{figure3}
  }
\end{figure}

We demonstrate these sequences by addressing individual $^{40}\text{Ca}^+$ ions confined in a microfabricated surface-electrode trap~\cite{Doret2012}; details of our surface trap setup can be found in~\cite{Vittorini2013}.  We use a 397~nm laser to Doppler cool and optically pump ions into the $|1\rangle = {^2}\text{S}_{1/2}\:(m_j = -1/2)$ state.  A narrow linewidth ($\gamma \sim 150$~Hz) 729~nm laser tuned to the $|1\rangle \rightarrow |0\rangle = {^2}\text{D}_{5/2}\:(m_j = -5/2)$ qubit transition is used to sideband cool the ion to $\leq 0.1$~phonons in all motional modes and to perform subsequent qubit rotations.  The 729~nm beam propagates parallel to the trap surface, intersecting the trap symmetry plane at a $45^\circ$ angle with a $w_0 = 44.2 \pm 0.8$~$\mu$m 1/$\e^2$ diameter waist.  Fast laser switching and phase control is achieved using an acousto-optic modulator driven by an amplified 16-bit direct-digital synthesizer with 20~ns timing resolution.  After applying a sequence of laser pulses, we measure the $|1\rangle$ state population using resonant ion fluorescence induced by the Doppler cooling laser.   

We verify our theoretical predictions for TASK1 sequences by measuring the qubit state-transfer for differing pulse areas, controlled by adjusting the timings of each laser pulse to scale the energy-time product by a constant multiple $\epsilon_j$.  The resulting propagation is thus equivalent to the evolution that would be experienced during an addressing error by neighboring ions over differing laser intensities.  Fig.~\ref{figure3}a and Fig.~\ref{figure3}b compare the measured response for pulse sequences applying $U_T = \exp(-\i \pi X / 2)$ and $U_T = \exp(-\i \pi X / 4)$ respectively to a target ion.  We observe that unwanted population inversion is suppressed when $\epsilon_j < 1$, as desired.  When $\epsilon_j \simeq 1$, corresponding to the pulse area experienced by the target ion, the observed state transfer is consistent with the expected gate.  We find excellent agreement between the calculated and measured response as a function of pulse area.

\begin{figure}
  \begin{center}
    \includegraphics[width=0.95\columnwidth]{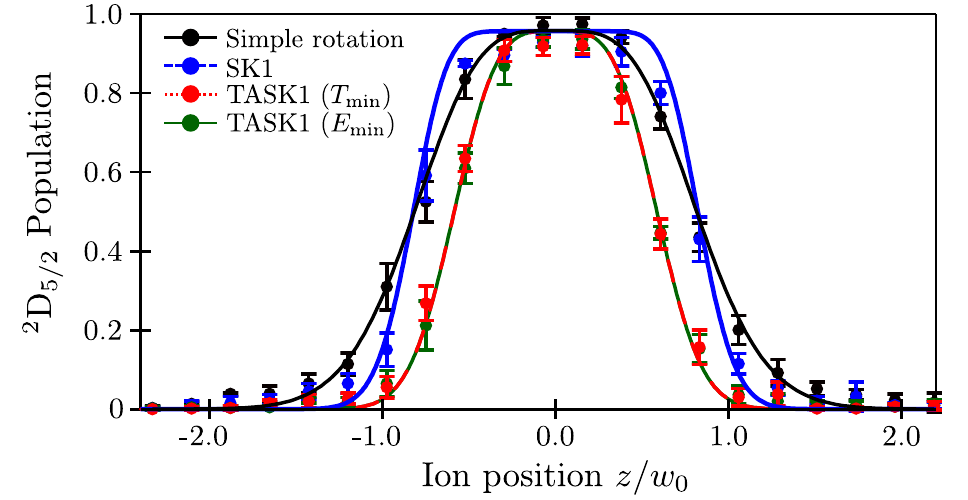}
  \end{center}
  \caption{Population inversion as a function of ion position for target $U_T = \exp(-\i \pi X/2)$ gates composed of simple rotations or TASK1 family pulses. The simple rotation maps out the intensity profile of the laser beam along the trap axis.  Narrowband sequences allow for an effective reduction in beam waist without changing the experimental setup.  Curves are theoretical predictions with a single, common adjustable parameter accounting for experimental qubit detection fidelity.
    \label{figure4}
  }
\end{figure}

In a second experiment we directly observe addressing error compensation by measuring the state transfer as a function of ion position relative to the center of the addressing beam (Fig.~\ref{figure4}).  We control ion position along the trap axis to better than $\pm 0.5$~$\mu$m by biasing a subset of 46 segmented DC trap electrodes~\cite{Doret2012}.  We find that narrowband sequences suppress unwanted rotations for ions placed far from the beam, while for ions aligned with the center of the laser profile we successfully execute the desired rotation.  

In conclusion, we introduced a Lie-algebraic transform method to produce narrowband sequences from other known sequences, frequently improving total pulse area and error suppression.  Using the technique, we developed the TASK1 family from transformations of SK1 and demonstrated their suitability for single-qubit addressing in an experiment with trapped ions.  Our transformation method is particularly well suited for narrowband sequences, where there is no desired unitary operation on the neighboring qubits.  Application of the mapping technique to other narrowband sequences, e.g. NB1 and the N2$j$ family \cite{Wimperis1994, Brown2004}, is straightforward.  Applying the technique to sequences where the errors occur on the addressed qubit, e.g. detuning and amplitude errors, should also be possible, however, there is additional complexity due to the control rotating the errors to a toggled frame \cite{Merrill2013}. 
In these cases Lie-algebraic maps cannot consist only of dilations but must also account for the frame transformation. 
Further, it is possible to concatenate these pulses with additional sequences that correct detuning errors.  These concatenated sequences should also assist with addressability concerns in systems with variable splitting frequencies, such as superconducting qubits~\cite{GambettaPRL2012}.

\begin{acknowledgments}
The authors thank J. M. Amini, A. M. Meier, and Kenton R. Brown for helpful discussions and critical review of the manuscript. GDV and SCD thank GTRI Shackelford Fellowships for graduate and postdoctoral support, respectively. This work was supported by the Office of the Director of National Intelligence - Intelligence Advanced Research Projects Activity through Army Research Office contract W911NF-10-1-0231 and Department of Interior contract D11PC20167. 
\end{acknowledgments}

\clearpage
\newpage
\onecolumngrid
\appendix

\section*{Supplementary Material: Table of TASK1 pulse sequences}
Here we provide explicit parameters for TASK1 pulse sequences that minimize the total pulse area or the residual infidelity.   Our paper describes rotation generators in terms of coordinates of vectors on a Lie algebra.  Here to aid in experimental implementation, we report each pulse in terms of a rotation angle and phase, $r_\ell = - \i \theta_\ell \sigma(\varphi_\ell) / 2$.  The reported sequences each generate a net rotation about the $X$ axis.  To shift the net rotation axis by an azimuthal angle $\varphi_T$, one adds $\varphi_T$ to the phase of each individual pulse in the sequence.  
 

\begin{table}[h!]
\begin{tabular}{c c c c c c c c c c c }
\hline \\[-1.0 em]
\hline \\[-0.7 em]
\multirow{2}{*}{Sequence}& \multirow{2}{*}{Net rotation} & \multirow{2}{*}{$\lambda_x$} & \multirow{2}{*}{$\lambda_y$} & $\theta_1$ & $\theta_2$ & $\theta_3$ & $\theta_4$ & $\theta_5$ & \multirow{2}{*}{Pulse area} & \multirow{2}{*}{Infidelity/$\epsilon_j^4$}\\
[-0.25 em] &  &  &  & $\varphi_1$ & $\varphi_2$ & $\varphi_3$ & $\varphi_4$ & $\varphi_5$ &  &  \\
\\[-0.7 em] \hline \\[-0.5 em]
\multirow{2}{*}{TASK1 ($T_\text{min}$)} & \multirow{2}{*}{$\pi/4$} & \multirow{2}{*}{0.2730} & \multirow{2}{*}{0.1828} & 0.6817 & 1.7155 & 1.3133 & 1.3133 & 0.6817 & \multirow{2}{*}{5.7055} & \multirow{2}{*}{0.0910} \\
[-0.1 em] &  &  &  & 1.5708 & 1.2177 & 3.5002 & 5.2184 & 4.7124 &  &  \\
[0.25 em]\multirow{2}{*}{} & \multirow{2}{*}{$\pi/2$} & \multirow{2}{*}{0.3988} & \multirow{2}{*}{0.2723} & 0.3013 & 2.5057 & 1.9404 & 1.9404 & 0.3013 & \multirow{2}{*}{6.9890} & \multirow{2}{*}{0.4308} \\
[-0.1 em] &  &  &  & 1.5708 & 1.2348 & 3.5075 & 5.2453 & 4.7124 &  &  \\
[0.25 em]\multirow{2}{*}{} & \multirow{2}{*}{$3\pi/4$} & \multirow{2}{*}{0.5000} & \multirow{2}{*}{0.3550} & 0.0000 & 3.1416 & 2.4898 & 2.4898 & 0.0000 & \multirow{2}{*}{8.1213} & \multirow{2}{*}{1.1510} \\
[-0.1 em] &  &  &  & 1.5708 & 1.2566 & 3.5101 & 5.2863 & 4.7124 &  &  \\
[0.25 em]\multirow{2}{*}{} & \multirow{2}{*}{$\pi$} & \multirow{2}{*}{0.5000} & \multirow{2}{*}{0.5000} & 0.0000 & 3.1416 & 3.1416 & 3.1416 & 0.0000 & \multirow{2}{*}{9.4248} & \multirow{2}{*}{2.2830} \\
[-0.1 em] &  &  &  & 1.5708 & 1.0472 & 3.1416 & 5.2360 & 4.7124 &  &  \\
[0.25 em]\multirow{2}{*}{} & \multirow{2}{*}{$5\pi/4$} & \multirow{2}{*}{0.5532} & \multirow{2}{*}{0.6227} & 0.1309 & 3.4758 & 3.8081 & 3.8081 & 0.1309 & \multirow{2}{*}{11.3539} & \multirow{2}{*}{4.3347} \\
[-0.1 em] &  &  &  & 4.7124 & 0.8958 & 2.9405 & 5.1343 & 1.5708 &  &  \\
[0.25 em]\multirow{2}{*}{} & \multirow{2}{*}{$3\pi/2$} & \multirow{2}{*}{0.6447} & \multirow{2}{*}{0.7135} & 0.3447 & 4.0507 & 4.3792 & 4.3792 & 0.3447 & \multirow{2}{*}{13.4984} & \multirow{2}{*}{7.7300} \\
[-0.1 em] &  &  &  & 4.7124 & 0.8251 & 2.8767 & 5.0567 & 1.5708 &  &  \\
[0.25 em]\multirow{2}{*}{} & \multirow{2}{*}{$7\pi/4$} & \multirow{2}{*}{0.7578} & \multirow{2}{*}{0.8246} & 0.5262 & 4.7613 & 5.0795 & 5.0795 & 0.5262 & \multirow{2}{*}{15.9728} & \multirow{2}{*}{14.2640} \\
[-0.1 em] &  &  &  & 4.7124 & 0.5860 & 2.6446 & 4.8106 & 1.5708 &  &  \\
[0.25 em]\multirow{2}{*}{} & \multirow{2}{*}{$2\pi$} & \multirow{2}{*}{1.0000} & \multirow{2}{*}{1.0000} & 0.0000 & 6.2832 & 6.2832 & 6.2832 & 0.0000 & \multirow{2}{*}{18.8496} & \multirow{2}{*}{36.5284} \\
[-0.1 em] &  &  &  & 4.7124 & 1.0472 & 3.1416 & 5.2360 & 1.5708 &  &  \\
[0.25 em]\multirow{2}{*}{TASK1 ($E_\text{min}$)} & \multirow{2}{*}{$\pi/4$} & \multirow{2}{*}{0.2226} & \multirow{2}{*}{0.2226} & 0.8001 & 1.3984 & 1.3984 & 1.3984 & 0.8001 & \multirow{2}{*}{5.7953} & \multirow{2}{*}{0.0896} \\
[-0.1 em] &  &  &  & 1.5708 & 1.0472 & 3.1416 & 5.2360 & 4.7124 &  &  \\
[0.25 em]\multirow{2}{*}{} & \multirow{2}{*}{$\pi/2$} & \multirow{2}{*}{0.3268} & \multirow{2}{*}{0.3268} & 0.4826 & 2.0534 & 2.0534 & 2.0534 & 0.4826 & \multirow{2}{*}{7.1255} & \multirow{2}{*}{0.4167} \\
[-0.1 em] &  &  &  & 1.5708 & 1.0472 & 3.1416 & 5.2360 & 4.7124 &  &  \\
[0.25 em]\multirow{2}{*}{} & \multirow{2}{*}{$3\pi/4$} & \multirow{2}{*}{0.4159} & \multirow{2}{*}{0.4159} & 0.2301 & 2.6134 & 2.6134 & 2.6134 & 0.2301 & \multirow{2}{*}{8.3002} & \multirow{2}{*}{1.0932} \\
[-0.1 em] &  &  &  & 1.5708 & 1.0472 & 3.1416 & 5.2360 & 4.7124 &  &  \\
[0.25 em]\multirow{2}{*}{} & \multirow{2}{*}{$\pi$} & \multirow{2}{*}{0.5000} & \multirow{2}{*}{0.5000} & 0.0000 & 3.1416 & 3.1416 & 3.1416 & 0.0000 & \multirow{2}{*}{9.4248} & \multirow{2}{*}{2.2830} \\
[-0.1 em] &  &  &  & 1.5708 & 1.0472 & 3.1416 & 5.2360 & 4.7124 &  &  \\
[0.25 em]\multirow{2}{*}{} & \multirow{2}{*}{$5\pi/4$} & \multirow{2}{*}{0.5841} & \multirow{2}{*}{0.5841} & 0.2301 & 3.6698 & 3.6698 & 3.6698 & 0.2301 & \multirow{2}{*}{11.4696} & \multirow{2}{*}{4.2510} \\
[-0.1 em] &  &  &  & 4.7124 & 1.0472 & 3.1416 & 5.2360 & 1.5708 &  &  \\
[0.25 em]\multirow{2}{*}{} & \multirow{2}{*}{$3\pi/2$} & \multirow{2}{*}{0.6732} & \multirow{2}{*}{0.6732} & 0.4826 & 4.2298 & 4.2298 & 4.2298 & 0.4826 & \multirow{2}{*}{13.6545} & \multirow{2}{*}{7.5020} \\
[-0.1 em] &  &  &  & 4.7124 & 1.0472 & 3.1416 & 5.2360 & 1.5708 &  &  \\
[0.25 em]\multirow{2}{*}{} & \multirow{2}{*}{$7\pi/4$} & \multirow{2}{*}{0.7774} & \multirow{2}{*}{0.7774} & 0.8001 & 4.8848 & 4.8848 & 4.8848 & 0.8001 & \multirow{2}{*}{16.2547} & \multirow{2}{*}{13.3445} \\
[-0.1 em] &  &  &  & 4.7124 & 1.0472 & 3.1416 & 5.2360 & 1.5708 &  &  \\
[0.25 em]\multirow{2}{*}{} & \multirow{2}{*}{$2\pi$} & \multirow{2}{*}{1.0000} & \multirow{2}{*}{1.0000} & 0.0000 & 6.2832 & 6.2832 & 6.2832 & 0.0000 & \multirow{2}{*}{18.8496} & \multirow{2}{*}{36.5284} \\
[-0.1 em] &  &  &  & 4.7124 & 1.0472 & 3.1416 & 5.2360 & 1.5708 &  &  \\
[0.25 em]\hline
\end{tabular}
\caption{Pulse sequences for TASK1 ($T_\text{min}$) and TASK1 ($E_\text{min}$). The rotation angle and phase for each pulse $r_\ell = - \i \theta_\ell \sigma(\varphi_\ell) / 2$ is  listed as well as the $\lambda_x$ and $\lambda_y$ used to scale the ASK1 sequence from the SK1 sequence. The last two columns show the total pulse area and the leading term of the infidelity. 
}
\end{table}

\end{document}